\documentclass[electronics,article,accept,pdftex,moreauthors]{Definitions/mdpi} 

\firstpage{1} 
\pubvolume{1}
\issuenum{1}
\articlenumber{0}
\pubyear{2025}
\copyrightyear{2025}
\externaleditor{Sergio Saponara and Abdussalam Elhanashi}
\datereceived{10 January 2025} 
\daterevised{30 January 2025} 
\dateaccepted{4 February 2025} 
\datepublished{ } 
\hreflink{https://doi.org/} 

\usepackage{longtable} 
\usepackage{makecell}

\Title{A Performance Analysis of You Only Look Once Models for Deployment on Constrained Computational Edge Devices in Drone Applications}

\TitleCitation{A Performance Analysis of You Only Look Once Models for Deployment on Constrained Computational Edge Devices in Drone Applications}

\Author{Lucas Rey\orcidA{}, 
 Ana M. Bernardos\orcidB{} *, Andrzej D. Dobrzycki\orcidC{},
 David Carramiñana\orcidD{}, Luca Bergesio\orcidE{}, Juan A. Besada\orcidF{}  \mbox{and José Ramón Casar\orcidG{}} 
 }

\AuthorNames{Lucas Rey, Ana M. Bernardos, Andrzej D. Dobrzycki, David Carramiñana, Luca Bergesio, Juan A. Besada and José Ramón Casar}

\AuthorCitation{
Rey, L.;
Bernardos, A.M.; Dobrzycki, A.D.; Carramiñana, D.; Bergesio, L.; Besada, J.A.; Casar, J.R.}

\address[1]{%
Information
 Processing and
Telecommunications Center, Universidad Politécnica de Madrid, 
ETSI Telecomunicación, Av. Complutense 30,
28040 Madrid, Spain; lucas.rey@upm.es (L.R.); \mbox{daniel.dobrzycki@upm.es (A.D.D.);} d.carraminana@upm.es (D.C.); luca.bergesio@upm.es (L.B.); juanalberto.besada@upm.es (J.A.B.); joseramon.casar@upm.es (J.R.C.)}

\corres{\hangafter=1 \hangindent=1.05em \hspace{-0.82em}Correspondence: anamaria.bernardos@upm.es}

\abstract{Advancements in embedded systems and Artificial Intelligence (AI) have enhanced the capabilities of Unmanned Aircraft Vehicles (UAVs) in computer vision. However, the integration of AI techniques o-nboard drones is constrained by their processing capabilities. In this sense, this study evaluates the deployment of object detection models (YOLOv8n and YOLOv8s) on both resource-constrained edge devices and cloud environments. The objective is to carry out a comparative performance analysis using a representative real-time UAV image processing pipeline. Specifically, the NVIDIA Jetson Orin Nano, Orin NX, and Raspberry Pi 5 (RPI5) devices have been tested to measure their detection accuracy, inference speed, and energy consumption, and the effects of post-training quantization (PTQ). The results show that YOLOv8n surpasses YOLOv8s in its inference speed, achieving 52 FPS on the Jetson Orin NX and 65 fps with INT8 quantization. Conversely, the RPI5 failed to satisfy the real-time processing needs in spite of its suitability for low-energy consumption applications. An analysis of both the cloud-based and edge-based end-to-end processing times showed that increased communication latencies hindered real-time applications, revealing trade-offs between edge (low latency) and cloud processing (quick processing). Overall, these findings contribute to providing recommendations and optimization strategies for the deployment of AI models \mbox{on UAVs.}}


\begin{document}

\setcounter{section}{0} 
\section{Introduction}
In recent years, the integration of unmanned aerial vehicles (UAVs), commonly referred to as drones, with machine learning techniques has significantly advanced the field of aerial robotics. Machine learning algorithms, particularly deep learning approaches, enable drone-based systems to process large volumes of data collected from onboard sensors. The incorporation of computer vision algorithms into drone technologies has enhanced their autonomous capabilities, allowing drones to detect and track objects with high accuracy. Additionally, these advancements support collaborative data fusion, where information from multiple drones and sensors is combined to improve situational awareness and inform decision-making processes. Such capabilities contribute to the effective management and coordination of drone operations in complex environments \cite{8969620, brown2023}, opening the door to possible applications such as vision-driven swarming \cite{8798720}.

While these intelligent capabilities offer significant advantages, embedding them into onboard or even edge systems remains a challenge. By integrating these algorithms directly into drones, they can operate autonomously without relying on constant communication with a central system. This autonomy is important in low-connectivity scenario applications, such as unsupervised maintenance flights in remote areas (e.g., oil pipelines or power pylons), defense operations, or emergency response situations where the connectivity may be limited \cite{profiling, 9693227}. In these cases, which also seek energy-efficient methods to avoid early battery drain, drones must be able to independently detect and track objects, and algorithms like YOLO (You Only Look Once) enhance their ability to do so effectively.

YOLOv8, as a state-of-the-art object detection algorithm, is well suited to these scenarios due to its multiple versions, with each optimized for different hardware constraints. For instance, versions such as YOLOv8-small (YOLOv8s) and YOLOv8-nano (YOLOv8n) contain fewer parameters, making them appropriate for deployment on the smaller, computationally constrained devices commonly used in edge environments. These versions trade off some accuracy for computational efficiency, allowing for real-time processing even on resource-constrained devices \cite{reis2024realtimeflyingobjectdetection}. Moreover, YOLOv8 demonstrates adaptability to varying input resolutions, which is important in applications such as drone operations, where the cameras can range from thermal imaging for rescue missions to high-definition cameras for other purposes \cite{ultralytics2023yolov8}. Additionally, its integration with tracking algorithms like DeepSORT supports object tracking in sequences of frames, which is valuable for scenarios that require dynamic monitoring, such as surveillance or search and rescue missions \cite{survey_apps}. YOLOv8 also supports multitasking by combining object detection with classification and segmentation in a single model, enabling it to address various computer vision tasks in controlled \mbox{settings \cite{ultralytics2023yolov8}.} As an evolving framework, using YOLOv8 as one of the latest versions of YOLO eases the transition to future developments of the algorithm.

Furthermore, techniques such as quantization can facilitate enhanced performance through reductions in the bit width, which refers to the number of bits used to represent data in computations. By reducing the bit width, the model performs calculations with fewer bits, thereby decreasing both the memory footprint and the computational load. This is relevant when running complex algorithms on devices with limited hardware capabilities, as it allows for more efficient use of the available resources without sacrificing the real-time performance. By applying quantization in combination with different fine-tuned YOLO models of varied sizes, we aim to derive guidelines for the optimal use of object detection algorithms like YOLOv8 in constrained environments. These insights may be useful for the future integration of such algorithms into distributed agents at the edge, where autonomy and performance are paramount \cite{profiling, DeepBench, app122010218}.

Given the challenges outlined, the primary objectives of this article are as follows:
\begin{enumerate}
  \item To comprehensively evaluate the performance of YOLOv8 models across different embedded platforms, including the Jetson Orin Nano, Jetson Orin NX, and Raspberry Pi 5, in order to determine their suitability for real-time object detection.
  \item To investigate the impact of various quantization techniques and floating-point precision models on the balance between the detection accuracy, processing speed, and power consumption, optimizing these trade-offs to ensure low-latency inference in edge computing applications.
  \item To validate the performance of the aforementioned devices in a representative deployment scenario for drone-based applications, which consists of integrating the models into a real drone image processing workflow. This integration also enables comparing edge and cloud solutions for object detection inference.
  \item To assess the specific limitations and potential of these platforms for real-time drone operations, identifying the constraints in terms of energy efficiency, algorithmic processing time, and scalability within real architectures, and to derive practical guidelines for selecting the appropriate models and devices for various deployment scenarios.

\end{enumerate}

The organization of this article is as follows. Section ~\ref{sec:sota} provides a comprehensive review of the state of the art, detailing advancements in deploying deep learning on energy-efficient devices into a distributed edge architecture and exploring quantization techniques in object detection models. Section ~\ref{sec:methodology} describes the methodology, including the evaluation setup, dataset creation, and the optimization of YOLO object detection models tailored to edge devices, alongside a detailed testing strategy. Section~\ref{sec:results} presents a rigorous evaluation of quantized YOLO models on resource-constrained devices, assessing their inference performance, accuracy degradation, and energy consumption. Section~\ref{sec:results2} expands on the performance in a realistic testbed, comparing edge and cloud solutions for object detection and discussing the practical implications. Finally, Section~\ref{sec:discussion} synthesizes the findings and proposes guidelines, while Section~\ref{sec:conclusions} concludes this paper, highlighting potential directions for future research, including the integration of digital twin simulations and enhanced edge--cloud collaboration for drone operations.

\section{The State of the Art}
This section is divided into two subsections. The first Section \ref{subsec:2.1}, Deploying Deep Learning on Energy-Efficient Devices, evaluates the performance and optimization techniques for deep learning models deployed on edge hardware, with a focus on energy efficiency and real-time processing. The second subsection, Section
 \ref{subsec:2.2}, Quantization Techniques in Object Detection Models, reviews strategies to optimize the computational and memory demands in object detection algorithms, analyzing their feasibility for edge devices in constrained environments.

\label{sec:sota}
\subsection{Deploying Deep Learning on Energy-Efficient Devices}
\label{subsec:2.1}
In recent years, significant research has focused on optimizing and evaluating the performance of deep learning models on edge devices, particularly for object detection tasks in resource-constrained environments. Various studies have benchmarked the hardware configurations to identify the optimal settings for achieving a balance between the inference speed, energy consumption, and real-time processing requirements. For instance, Baller et al. \cite{DeepBench} evaluated a range of edge devices, including the NVIDIA Jetson Nano, and identified the optimal configurations in terms of the inference speed and energy consumption across various deep learning frameworks and classification models. Such evaluations are crucial for understanding the limitations and potential of hardware in real-time scenarios.

Building on hardware-specific optimizations, Li et al. introduced \cite{ABM-SpConv-SIMD}, an optimization framework for CNN inference on edge devices for application in an IoT context. This framework combines offline pruning and quantization techniques with runtime optimizations, such as multiplication reduction, data layout adjustments, and data parallelization. The experimental results showed performance improvements of 1.96$\times$ in high-end devices and 1.73$\times$ in low-end devices in improving their edge device efficiency using advanced model compression techniques. Similarly, Ferraz et al. \cite{Ferraz} benchmarked CNN inference on edge devices, such as the NVIDIA Jetson AGX Xavier and the Movidius Neural Stick, noting considerable improvements in the inference time and power efficiency. 

Recent studies also emphasize the use of distributed and cooperative processing to enhance edge device performance. Hou et al. \cite{DistrEdge:} introduced DistrEdge, a method designed to enhance CNN inference across distributed edge devices by employing deep reinforcement learning. This approach distributes inference tasks among multiple devices, optimizing the computational load and significantly improving the overall inference speed. DistrEdge dynamically adapts to the heterogeneity of devices and varying network conditions, ensuring efficient load balancing and reduced latency. This adaptability makes it particularly effective in real-time processing scenarios. In another cooperative approach, Liang et al. \cite{liang2022} presented Edge YOLO, leveraging edge--cloud cooperation for real-time object detection in autonomous driving, underscoring the benefits of a distributed architecture in complex, time-sensitive tasks.

In the realm of object detection, Shin and Kim \cite{shin2022} explored the performance of YOLO object detection models on NVIDIA Jetson devices, highlighting the advantages of using TensorFlow-TensorRT (TF-TRT) to optimize real-time inference. They also found that while TF-Lite is effective for mobile, it fails to utilize the GPU resources on the Jetson, highlighting the need to select appropriate frameworks based on the hardware and application needs. Xiong et al. \cite{xiong2023} further explored YOLO models by optimizing YOLOv5s for UAV image detection on the Jetson Xavier NX, achieving notable improvements in FPS and model compression. Similarly, Hu et al. \cite{hu2023} proposed an improved YOLO-Tiny-attention algorithm for fault detection on wind turbine blades, which was successfully deployed on the same hardware with increased detection precision. These findings highlight the adaptability of YOLO-based models in different edge applications.

Table \ref{tab:comparison} summarizes the key findings from these studies, highlighting the energy consumption and inference performance across different edge devices and models.
\begin{table}[H]
\tablesize{\small}
\caption{Comparison of energy efficiency and inference latency across reviewed studies. \label{tab:comparison}}

\begin{adjustwidth}{-\extralength}{0cm}
\centering
\begin{tabularx}{\fulllength}{CCCCCC}
    \toprule
    \textbf{Authors} & \textbf{Device} & \textbf{Model} & \textbf{Metrics} & \textbf{Application Area} & \textbf{Key Findings} \\ 
    \midrule
    Baller et al. 2021 \cite{DeepBench} & NVIDIA Jetson Nano & MobileNetV2, DNNs & Inference speed, accuracy, energy consumption & General classification & Optimized frameworks needed to improve inference time and energy use in Jetson Nano. \\ 
    \midrule
    Li et al. 2023 \cite{ABM-SpConv-SIMD} & NVIDIA Jetson Nano, Raspberry Pi & MobileNetV2, VGG16 & Inference speed, energy consumption & IoT applications & Jetson Nano is faster and more energy-efficient than Raspberry Pi, but neither meets real-time needs for complex tasks. \\ 
    \midrule
    Ferraz et al. 2023 \cite{Ferraz} & NVIDIA Jetson AGX Xavier, Movidius Neural Stick & SqueezeNet, CNNs & Inference speed, energy consumption & Edge devices & Pruning and quantization improve energy efficiency and performance on AGX Xavier. \\ 
    \midrule
    Hou et al. 2022 \cite{DistrEdge:} & NVIDIA Jetson Nano, TX2, Xavier & CNN & Distributed inference performance, speedup & Edge devices & Distributed inference framework (DistrEdge) with reinforcement learning improves real-time performance. \\

\bottomrule
\end{tabularx}\end{adjustwidth}
\end{table}

\begin{table}[H]\ContinuedFloat
\caption{{\em Cont.}}\label{tab:comparison}

\begin{adjustwidth}{-\extralength}{0cm}
\centering
\begin{tabularx}{\fulllength}{CCCCCC}
    \toprule
    \textbf{Authors} & \textbf{Device} & \textbf{Model} & \textbf{Metrics} & \textbf{Application Area} & \textbf{Key Findings} \\ 

 \midrule
    Liang et al. 2022 \cite{liang2022} & Edge--cloud, NVIDIA Jetson & Edge YOLO & Real-time performance, object detection accuracy & Autonomous vehicles & Pruning and feature compression improve efficiency in edge-cloud systems, enhancing autonomous vehicle operations. \\ 
    \midrule
    Shin and Kim 2022 \cite{shin2022} & NVIDIA Jetson AGX Xavier & YOLOv4-Native, YOLOv4-Tiny & Latency, energy consumption & Real-time object detection & TF-TRT and TensorRT optimizations improve real-time inference and reduce energy use on Jetson AGX Xavier. \\ 
    \midrule
    Xiong et al. 2023 \cite{xiong2023} & NVIDIA Jetson Xavier NX & GCGE-YOLO & mAP, FPS, energy consumption & UAV image detection & Uses GhostNet and coordinate attention to reduce computational load and improve object detection accuracy. \\ 
    \midrule
    Hu et al. 2023 \cite{hu2023} & NVIDIA Jetson Xavier NX, Jetson Nano & YOLO-Tiny-attention & FPS, detection precision & Wind turbine fault detection & Attention mechanisms improve fault detection accuracy on low-power edge devices, supporting real-time monitoring. \\ 
    \bottomrule
\end{tabularx}
\end{adjustwidth}
\end{table}

While many studies focus on evaluating deep learning algorithms on individual edge devices, such as the NVIDIA Jetson Nano or AGX Xavier, there is growing recognition of the need to explore more complex architectures. These include distributed computing systems and edge--cloud cooperation, which are gaining significance for handling sophisticated tasks in real-time applications. Several studies, such as Hou et al. \cite{DistrEdge:} and Liang et al. \cite{liang2022}, have begun to explore the potential of distributed architectures to overcome the limitations of individual devices, demonstrating the benefits of dynamic workload balancing and feature compression in real-time object detection tasks. 

Distributed edge architectures, when integrated with tactical cloud technology and 5G networks, enable real-time data flow among edge devices \cite{xu2022}. This integration enhances the performance of autonomous drones by supporting efficient coordination and rapid decision-making. By combining local data processing with continual connectivity, such architectures improve the operational efficiency and autonomy in dynamic \mbox{environments \cite{cai2022}.} In addition, the network slice feature segments the network into specific virtual environments to enhance the resource allocation and service quality per application \cite{DBLP:journals/corr/abs-2112-07048}. This ensures that in drone systems, each mission or drone group receives a tailored ``slice'' with low latency and high reliability, important for enabling efficient operations.

By integrating these advanced communication frameworks with quantization techniques \cite{Jiang2022Post, przewlocka2022, al-hamid2023}, researchers can improve the energy efficiency and performance in edge systems. This combination of tactical cloud, 5G, and model quantization addresses the specific challenges in distributed architectures, such as latency reduction and resource optimization. These improvements support the development of more autonomous and efficient systems capable of operating effectively in energy-constrained real-world scenarios.

\subsection{Quantization Techniques in Object Detection Models}
\label{subsec:2.2}
Quantization techniques are valuable for optimizing deep learning models for deployment in edge devices, improving the inference speed and energy efficiency, especially in real-time applications such as autonomous drone navigation and object detection. Common techniques include post-training quantization (PTQ), Quantization-Aware Training (QAT), and mixed-precision quantization (MPQ), all of which enhance the energy efficiency in reducing the computational complexity while maintaining the model's performance.

PTQ is a technique that reduces the precision of a trained model to formats such as 8-bit integers by converting the weights and activations from floating-point representations. One of its primary advantages is that it reduces the size of the model and increases the inference speed without requiring retraining, preserving the original training process. For instance, Przewlocka-Rus et al. \cite{przewlocka2022} demonstrated that PTQ applied to object detection models on the NVIDIA Jetson Nano decreased the inference times by up to 35\% with a less than 2\% degradation in accuracy. Similarly, Jiang et al. \cite{Jiang2022Post} evaluated various PTQ approaches, including affine, logarithmic, and dynamic quantization, and found that these methods effectively minimized the size of models while maintaining high accuracy with minimal losses.

QAT integrates quantization into the training process, allowing the model to learn to minimize quantization errors. This method typically results in higher accuracy than PTQ, especially for complex models and tasks. Gupta and Asthana \cite{Gupta2023Reducing} applied QAT to object detection models such as YOLO, achieving a 1.8\% improvement in mAP compared to PTQ, with reductions in the model size and inference time.

MPQ uses different precisions for various parts of the model, balancing trade-offs between size, speed, and accuracy. Critical layers use higher precision, while less critical layers use lower precision. Park et al. \cite{zhou2023} explored mixed-precision quantization in object tracking models, achieving energy consumption reductions of up to 45\% and inference time improvements of 30\% on edge devices like the Jetson Nano and Raspberry Pi. Furthermore, Al-Hamid and Kim \cite{al-hamid2023} proposed a Unified Scaling-Based Pure-Integer Quantization (USPIQ) method that reduced the on-chip memory by 75\% with only a 0.61\% loss in the mAP while achieving a 2.84x speedup in the inference time compared to traditional methods.

These quantization approaches improve the energy and space efficiency in edge device deployments, making the implementation of sophisticated object detection and tracking models viable. However, the effects of these techniques on the latest YOLO models and their integration into real-time drone applications are still being researched. Studies on deploying YOLO models in different edge--cloud or embedded edge implementations are limited \cite{Jiang2022Post, przewlocka2022, zhou2023, al-hamid2023}.

Despite significant advances in the research on object detection at the edge, several gaps remain. Multiple studies have assessed edge devices in isolation without results on the latency in the interaction between the hardware and software processes.~

To address these gaps, this study focuses on YOLOv8 models applied to edge environments, emphasizing their accuracy, inference speed, and energy consumption. By incorporating quantization techniques, this research offers new insights into the potential of quantization approaches to optimize YOLOv8 for resource-limited edge scenarios. The findings deliver practical recommendations for selecting the configurations in power-constrained environments, with a particular focus on the Jetson Orin Nano, Jetson Orin NX, and Raspberry Pi 5 as innovative edge devices.

\section{Methodology}
\label{sec:methodology}

In this study, three edge computing devices are used to evaluate the performance of YOLOv8 models and their quantized versions: the Raspberry Pi 5, the Jetson Orin Nano, and the Jetson Orin NX. These devices were chosen due to their different hardware capabilities, allowing for an evaluation of object detection tasks under various processing constraints.~\tablename~\ref{tab:hardware_specs} provides an overview of the key hardware specifications for \mbox{each device.}

\begin{table}[H]
\caption{Comparison of hardware characteristics of the devices used in this study.\label{tab:hardware_specs}}
\begin{tabularx}{\textwidth}{CCCC}
    \toprule
    \textbf{Feature} & \textbf{Raspberry Pi 5} & \textbf{Jetson Orin Nano} & \textbf{Jetson Orin NX} \\
    \midrule
    CPU 
 & 4-core ARM Cortex-A76 at \mbox{2.4 GHz} & 6-core ARM Cortex-A78AE & 8-core ARM Cortex-A78AE \\\midrule
    GPU & VideoCore VII & 1024-core Ampere + 32 Tensor Cores & 2048-core Ampere + 64 Tensor Cores \\\midrule
    RAM & 8 GB & 8 GB & 16 GB \\\midrule
    Memory Bandwidth & Limited & Moderate & High \\\midrule
    Typical Power Consumption & 5–7 W & 7–15 W & 10–25 W \\\midrule
    TensorRT Support & No & Yes & Yes \\\midrule
    Dimensions & 85.6 $\times$ 56.5 mm & 100 $\times$ 69.6 mm & 100 $\times$ 87 mm \\
    \bottomrule
\end{tabularx}
\end{table}

These devices will be used to evaluate the performance of YOLOv8 models and their quantized versions under uniform experimental conditions. The Raspberry Pi 5, a CPU-centric device, was selected to assess the capabilities of a modern single-board computer without a dedicated GPU in object detection tasks. In contrast, the Jetson Orin Nano and Jetson Orin NX, which are GPU-powered platforms, were chosen for their high-performance processing capabilities. Together, these devices represent a range of edge computing architectures, enabling an evaluation of the inference speed, accuracy, and energy consumption across diverse hardware configurations.

In order to carry out these evaluations, it is necessary to first have an object detection model. To this end, Section \ref{sec:modelTraining} describes the process of creating such a model, including the generation of a dedicated dataset in \ref{sec:dataset}, model selection and training in \ref{sec:modelSelection}, and model optimization in \ref{sec:modelOptimization}. From the model, two types of experiments \mbox{are developed:} 
\begin{itemize}
    \item Experiments that deploy the model in isolation on each of the previous hardware platforms. These experiments aim to evaluate the inference speed and power consumption, for which the results are detailed in Section \ref{sec:ResultsEdge}.
    \item Experiments that integrate the object detection model into a typical drone image processing pipeline. As described in Section \ref{sec:results2}, these experiments seek to compare the capabilities of edge devices against a cloud deployment under a \mbox{representative scenario.}
\end{itemize}

In these experiments, a set of metrics are used to evaluate the performance of the models. These metrics are described in Section \ref{sec:Metrics}.

\subsection{Object Detection Model Selection and Training}
\label{sec:modelTraining}

\subsubsection{Dataset Creation}
\label{sec:dataset}
\begin{figure}[H]
\includegraphics[width=10.5 cm]{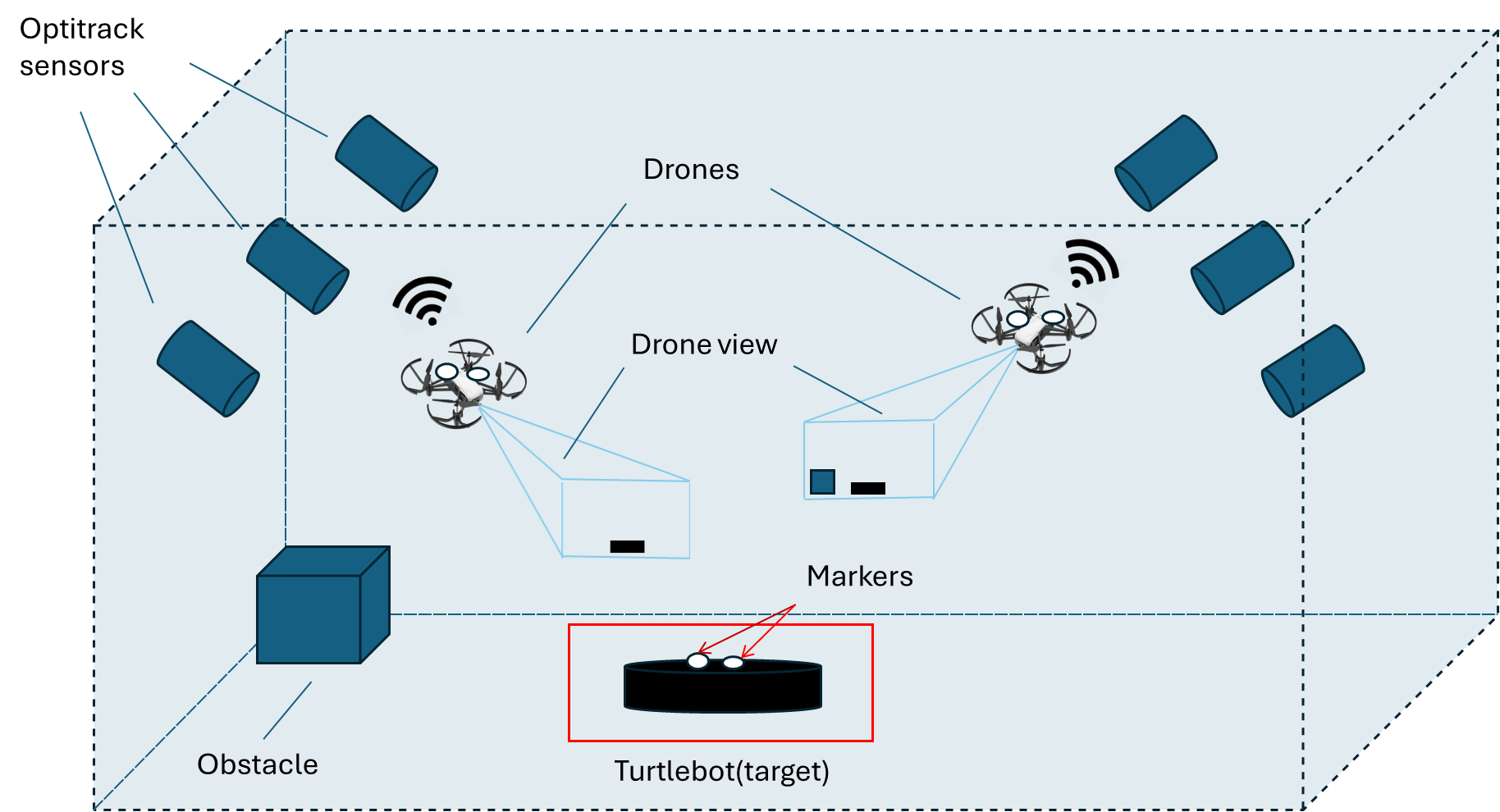}
\caption{Diagram of the indoor controlled flight environment equipped with OptiTrack sensors, drones with access points, and a TurtleBot target on the ground.}
\label{fig:drone_room}
\end{figure}

To train the evaluated object detection models, a dataset of annotated images was created. This dataset was collected within the indoor testbed described by the authors in \cite{David_Carramiñana_IoT} and depicted in Figure \ref{fig:drone_room}. The test setup consists of an indoor room with varied furniture elements to introduce a complex environment. Within the room, a fleet of drones (DJI Tello) can be controlled, together with a fleet of ground robots (TurtleBot) that serve as the targets (i.e., objects) to be detected from the aerial images. The experimentation platform allows you to easily capture frames from the drone's onboard cameras. Also, positioning information can be simultaneously gathered thanks to an OptiTrack system (i.e., a commercial high-precision positioning system based on high-speed motion capture cameras). Although this former capability was not exploited, it could be useful in the future to enable automatic target labeling in captured images. 

\begin{figure}[H]
\includegraphics[width=10.5 cm]{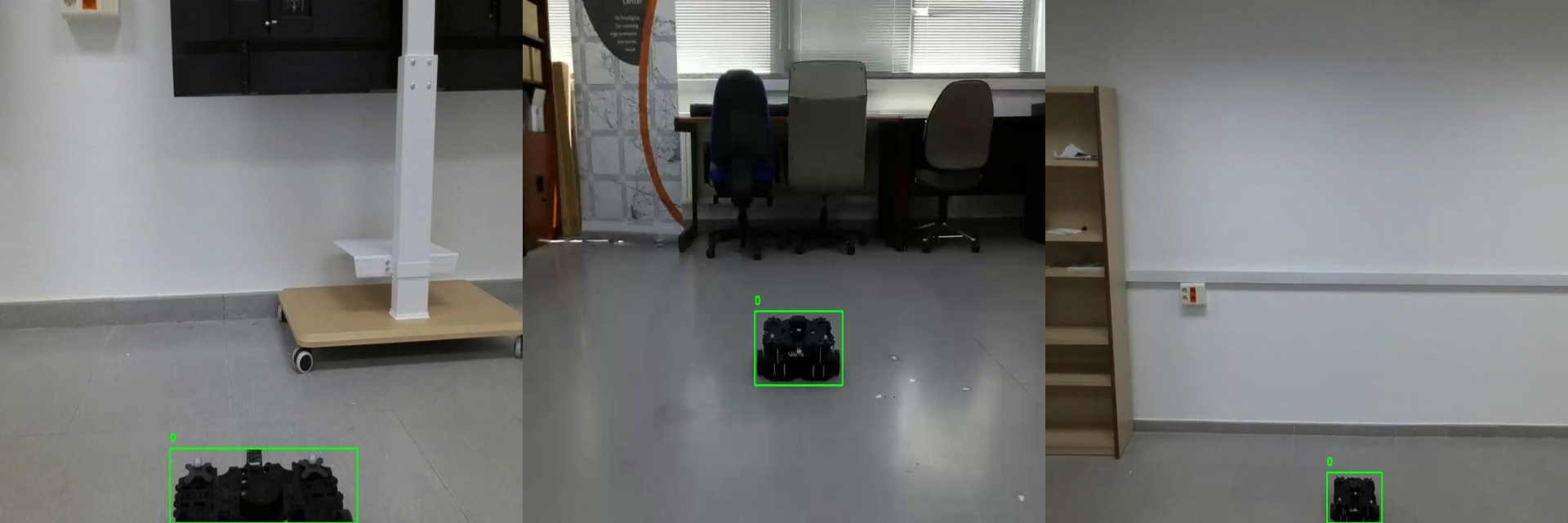}
\caption{Examples of different dataset images where the object to be detected has been manually annotated. Images are captured from different angles and heights and with different background environments.}
\label{fig:dataset}
\end{figure}

The dataset is composed of 6000 aerial images (640 $\times$ 640 resolution) with the target in different positions relative to the drone's field of view.  In order to ensure that the dataset includes a range of angles and perspectives, the drones were  flown following various straight-line and diagonal trajectories relative to the target position. Also, altitude changes were also introduced to capture different viewing angles, with certain trajectories varying from 1.5 m down to 0.5. After image gathering, the dataset was manually annotated, as demonstrated by the example images in Figure \ref{fig:dataset}. In general, this dataset serves as the foundation for training and evaluating object detection models, with annotated images integrated into the YOLO object detection pipeline.

\subsubsection{Selection and Training of Object Detection Models}
\label{sec:modelSelection}

For the selection of the model for this study, it was necessary to evaluate different versions of the YOLO architecture in terms of their performance and efficiency. YOLOv9 was considered a potential candidate; however, YOLOv8 demonstrated superior optimization for real-time performance and latency reduction. Moreover, YOLOv8 was elected as the most stable version during the experimental advance of this research.

To align with the objective of deploying models on devices with limited computational resources further, this study specifically focused on the `small' (YOLOv8s) and `nano' (YOLOv8n) versions of YOLOv8.  These versions are designed for computationally constrained devices by reducing the depth and number of parameters, making them suitable for various edge deployment scenarios. YOLOv8n, the smallest and fastest version of the model in this family, is optimized for devices with strict resource limitations. Its architecture prioritizes the inference speed by using fewer convolutional layers and simplifying the feature extraction stages.

In contrast, YOLOv8s includes more convolutional layers and feature extraction steps, improving the detection accuracy while maintaining computational efficiency. This model provides a balanced approach, offering a slight increase in the computational demand for better precision, making it suitable for scenarios requiring both fast response times and high accuracy.~Both models share features such as anchor-free detection and enhanced feature pyramid networks, supporting their adaptability to diverse edge \mbox{deployment requirements.}

The models were trained using the AdamW optimizer, chosen for its ability to efficiently manage weight updates while applying decoupled weight decay. A learning rate of 0.002 was selected, ensuring fast convergence while avoiding divergence or overfitting, and a weight decay of 0.0005 was applied to most weights, excluding biases. This regularization approach encourages smaller weights, making the model more robust to noise, while excluding biases avoids penalizing parameters critical to feature representation. The momentum was set to 0.9 as the standard value to smooth optimization trajectories and ensure stable weight updates over training iterations. Training was conducted over 100 epochs with a batch size of 16, which was limited by the GPU hardware constraints of the machines used for training, balancing the computational cost and generalization performance. Early stopping with a patience of 10 epochs prevented overfitting by halting the training when no significant validation improvements were observed, saving computational resources. Data augmentation techniques, including mosaic and random erasing, were employed to enhance the diversity of the training samples. Mosaic augmentation combined multiple images into a single input, enabling the model to learn from varying object positions and scales, which is particularly effective for detection tasks. Random erasing increased the robustness by encouraging the model to focus on discriminative features rather than memorizing specific regions. During validation, predictions were made with a confidence threshold of 0.5, ensuring only reliable detections were considered. An IoU threshold of 0.7 defined the overlap criterion for successful detections, achieving a balance between precision \mbox{and recall.}

\subsubsection{Optimization Techniques for Edge Deployment}
\label{sec:modelOptimization}

To optimize the original models (small and nano) for edge devices, several techniques were employed, including reducing the precision to FP16 and applying INT8 quantization.

FP16, or Half-Precision Floating Point, reduces the bit precision of floating-point numbers from 32 bits (FP32) to 16 bits. This precision reduction, often implemented using tools such as TensorRT during model export, enhances the performance by decreasing the computational load and memory usage. Despite reducing the precision, FP16 offers inference speed improvements without significantly impacting the accuracy.

INT8 quantization optimizes models by converting weights and activations from floating-point precision into 8-bit integers. This process, often applied through post-training quantization (PTQ), significantly reduces the computational demand and improves the energy efficiency and inference speed, making it suitable for resource-constrained edge devices. However, unlike FP16, which typically preserves the model accuracy, INT8 quantization can lead to more noticeable accuracy degradation, particularly in complex tasks or when it is applied to models with high parameter sensitivity. 

In this work, TensorRT and NCNN were selected as the primary frameworks for optimizing and deploying the YOLOv8 models on the Jetson devices and the Raspberry Pi 5, respectively. TensorRT, deeply integrated with NVIDIA hardware, improves the model performance through layer fusion and memory optimization, enabling efficient inference with support for FP32, FP16, and INT8 quantization. ONNX (Open Neural Network Exchange) served as an intermediary format, facilitating precision reduction and compatibility during the model's conversion to TensorRT.
For ARM-based architectures like the Raspberry Pi, NCNN was chosen due to its lightweight design and low-latency inference capabilities, which are particularly effective on mobile and edge devices. While other frameworks, such as CoreML and TorchScript, were considered, they were deemed less suitable for this project’s goals. CoreML, tailored to iOS, was excluded due to the platform's irrelevance, and TorchScript did not align as closely with the optimization requirements as TensorRT and NCNN.

By tailoring the frameworks and quantization techniques to each deployment environment, the optimization strategy leveraged hardware-specific strengths. For the Jetson Orin Nano and NX, TensorRT was used to apply three quantization levels—FP32, FP16, and INT8—to establish a comparison. On the Raspberry Pi 5, only the FP32 format, as the original format for YOLO, was utilized via the NCNN framework.

\subsection{The Evaluation Strategy and Metrics}
\label{sec:Metrics}

As previously introduced, the experiments for measuring the performance of the YOLO models were performed in two stages to measure the set of metrics described in this section, shown on Table \ref{tab:performance_metrics_summary}.
Initially, in a controlled and isolated setting, we measured the latency, throughput, and energy efficiency to balance the inference speed and accuracy against edge device constraints. The experiments carried out, whose results are described in Section \ref{sec:results}, included the following:

\begin{itemize}
    \item \textbf{Inference Time Measurement
}: This experiment assessed the computational efficiency of each model in isolation by measuring the time to perform inferences on 5000 images at 640 $\times$ 640 resolution. The average inference time was recorded, and the MIS (FPS) metric was derived to indicate the processing capacity.

    \item \textbf{Continuous Inference Evaluation (Inferences Per Minute, IPM)}: In this experiment, each device performed inferences continuously over 60 s, accounting for all of the latency factors, including model loading, network overhead, and processing delays. The IPM metric provided insight into the system's throughput under real-world conditions.

    \item \textbf{Energy Consumption Analysis}: Real-time power usage was monitored using \texttt{tegrastats} for Jetson devices and \texttt{vcgencmd pmic\_read\_adc} for the Raspberry Pi 5. Power data, averaged over the duration, were used to calculate the energy consumption (W·s) and energy per inference (J/inference), indicating the energy efficiency necessary for battery-powered deployments.

\end{itemize}

Secondly, trials were performed in real-world scenarios to assess the performance under actual conditions. This systematic testing provides a comprehensive grasp of the abilities of the model, from isolated evaluations to complete operational contexts. In the latter case, the following test was performed for the results discussed in Section \ref{sec:results2}:

\begin{itemize}

    \item \textbf{Throughput and Latency Assessment}: The throughput times were measured in both edge and cloud environments using a timestamp-based approach involving a video server, a processing agent, and a prediction client. Timestamps captured the communication times and processing delays, with the Round-Trip Time (RTT) and prediction time for cloud processing allowing for precise comparisons of the bottlenecks across setups.
\end{itemize}

Table \ref{tab:performance_metrics_summary} provides a concise overview of the key metrics used throughout the previous experiments for evaluating the model efficiency, energy usage, and system responsiveness. These metrics were selected to capture distinct aspects of the system performance for real-time processing on edge devices. MIS and IPM measure the inference speed and hardware throughput, respectively, while energy consumption and EPI focus on energy efficiency, needed not only for the design of battery-powered operations but also for maximizing the relative energy utilization based on the relative energy wasted per inference. Together, these metrics provide a comprehensive framework for analyzing and comparing models across different devices and configurations.

In addition to computational metrics, predictive accuracy was assessed using mAP50 and mAP50-95, which evaluate the detection quality across quantization levels and settings. These metrics ensure that the whole evaluation captures both efficiency and reliability.

\begin{table}[H]
\caption{Summary of the performance metrics with their purpose and formulas. \label{tab:performance_metrics_summary}}
\begin{adjustwidth}{-\extralength}{0cm}
\centering
\renewcommand{\tabcolsep}{0.2cm}
\begin{tabularx}{\fulllength}{CCC}
    \toprule
    \textbf{Metric} & \textbf{Purpose} & \textbf{Formula} \\
    \midrule
    \multirow{2}{*}{Model Inference Speed or MIS (FPS)} 
    & Measures the speed of model inference in frames per second (FPS), reflecting the real-time processing capability. This measure will only measure the capability of the algorithm in an isolated way, without counting extra processing tasks that produce latency.
    & $\text{MIS} = \frac{\text{1}}{\text{Mean Inference Time (s)}}$ \\
    \midrule
    \multirow{2}{*}{Inferences Per Minute or IPM} 
    & Calculates the total inferences in 60 s, representing the overall device throughput. IPM assesses the entire process within the device, including I/O operations, to evaluate the hardware performance comprehensively. It measures the efficiency of the program within the isolated environment of the device.
    & IPM = Total inferences in one minute \\
    \midrule
    \multirow{2}{*}{Energy Consumption or E (W·s)} 
    & Indicates the total energy used during model inference, assessing the power consumption. Measured using device-specific tools (\texttt{tegrastats}
 for the Jetson and \texttt{vcgencmd pmic\_read\_adc} for the Raspberry Pi), allowing for energy estimation through the average power consumption.
    & $E = \bar{P} \cdot (t_2 - t_1)$ \\

\bottomrule
\end{tabularx}\end{adjustwidth}
\end{table}

\begin{table}[H]\ContinuedFloat

\caption{{\em Cont.}}
\label{tab:performance_metrics_summary}
\begin{adjustwidth}{-\extralength}{0cm}
\centering
\renewcommand{\tabcolsep}{0.2cm}
\begin{tabularx}{\fulllength}{CCC}
    \toprule
    \textbf{Metric} & \textbf{Purpose} & \textbf{Formula} \\

 \midrule
    {Energy Per Inference or EPI \linebreak (J/inference)} 
    & Lower EPI values signify a more efficient use of energy per prediction, providing insight into how effectively the model utilizes the available power for inference tasks. As a relative metric, EPI advances our understanding of energy efficiency comparisons across different models and systems. 
    & $\text{EPI} = \frac{\text{Mean Power (W)} \times 60 \text{ s}}{\text{IPM}}$ \\
    \midrule
    \multirow{2}{*}{Round-Trip Time or RTT (ms)} 
    & Measures the end-to-end time for a complete inference cycle, including the network latency and processing time, providing a comprehensive assessment of the system's whole architecture's real-time performance.
    & $\text{RTT} = \text{Client Reception time} - \text{Request sent time}$ \\
    \midrule
   {Throughput of the System\linebreak(Inferences/s)} 
    & Represents the system's capacity for continuous inferences, derived from the inverse of the RTT. Indicates the efficiency of the whole architecture in handling continuous inference requests over time.
    & $\text{Throughput} = \frac{1}{\text{Average RTT (s)}}$ \\
    \bottomrule
\end{tabularx}
\end{adjustwidth}
\end{table}

This experimental setup offers a robust framework for evaluating YOLOv8 models on edge devices. By systematically progressing from isolated testing to real-world deployment, it combines deployment techniques with a detailed performance analysis. This approach supports energy-efficient model optimization tailored to diverse hardware scenarios, ensuring suitability for real-time, low-powered applications and aligning with the goal of delivering practical evaluation strategies for edge-based implementations.

\section{Performance Evaluation of Quantized YOLO over Resource-Constrained Devices}
\label{sec:ResultsEdge}
\begin{figure}[H]
\begin{adjustwidth}{-\extralength}{0cm}
\centering
\includegraphics[width=15.5cm]{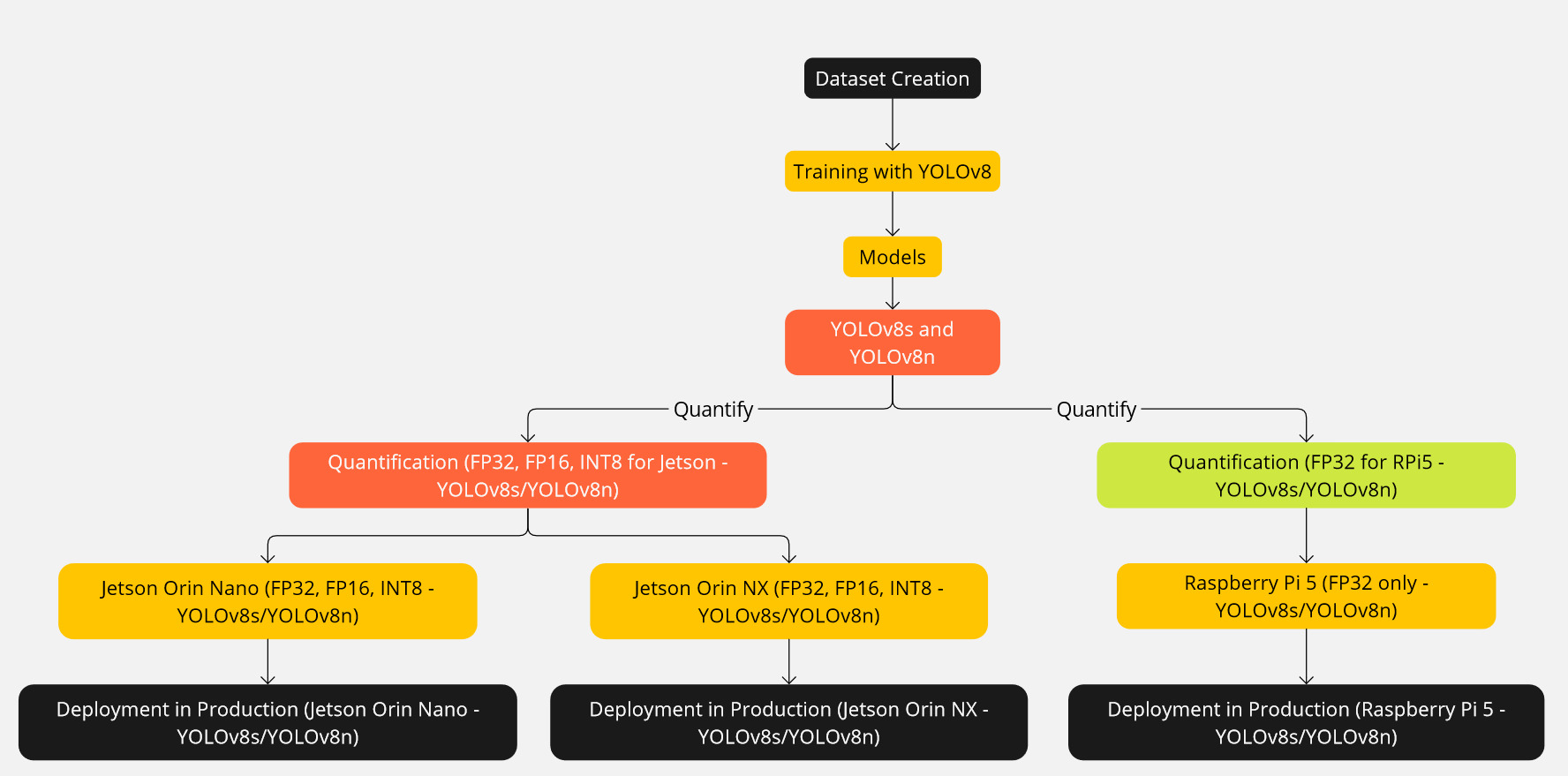}
\end{adjustwidth}
\caption{Quantization and deployment process of YOLOv8 models on the Jetson Orin Nano, Jetson Orin NX, and Raspberry Pi 5.}
\label{fig:quantization}
\end{figure}

This section provides a comprehensive evaluation of the performance of quantized YOLOv8 models on different resource-constrained devices, emphasizing the effects of the quantization techniques and hardware constraints on the model's efficiency and accuracy. \mbox{\figurename~\ref{fig:quantization}} depicts the structured workflow, from dataset creation to model training, followed by quantization and deployment. It showcases the distinct quantization formats (FP32, FP16, and INT8) applied to various devices like the Jetson Orin Nano, Jetson Orin NX, and Raspberry Pi 5, along with their associated production deployment. In particular, FP32 represents the default training format used for this type of model, which is sometimes also treated as the ``original'' model in the following sections. This visual workflow supports comprehension of the process and identifies the specific models examined in the experiments.

\subsection{The Inference Performance} \label{sec:results}

\begin{figure}[H]
\begin{adjustwidth}{-\extralength}{0cm}
\centering
\includegraphics[width=15.5cm]{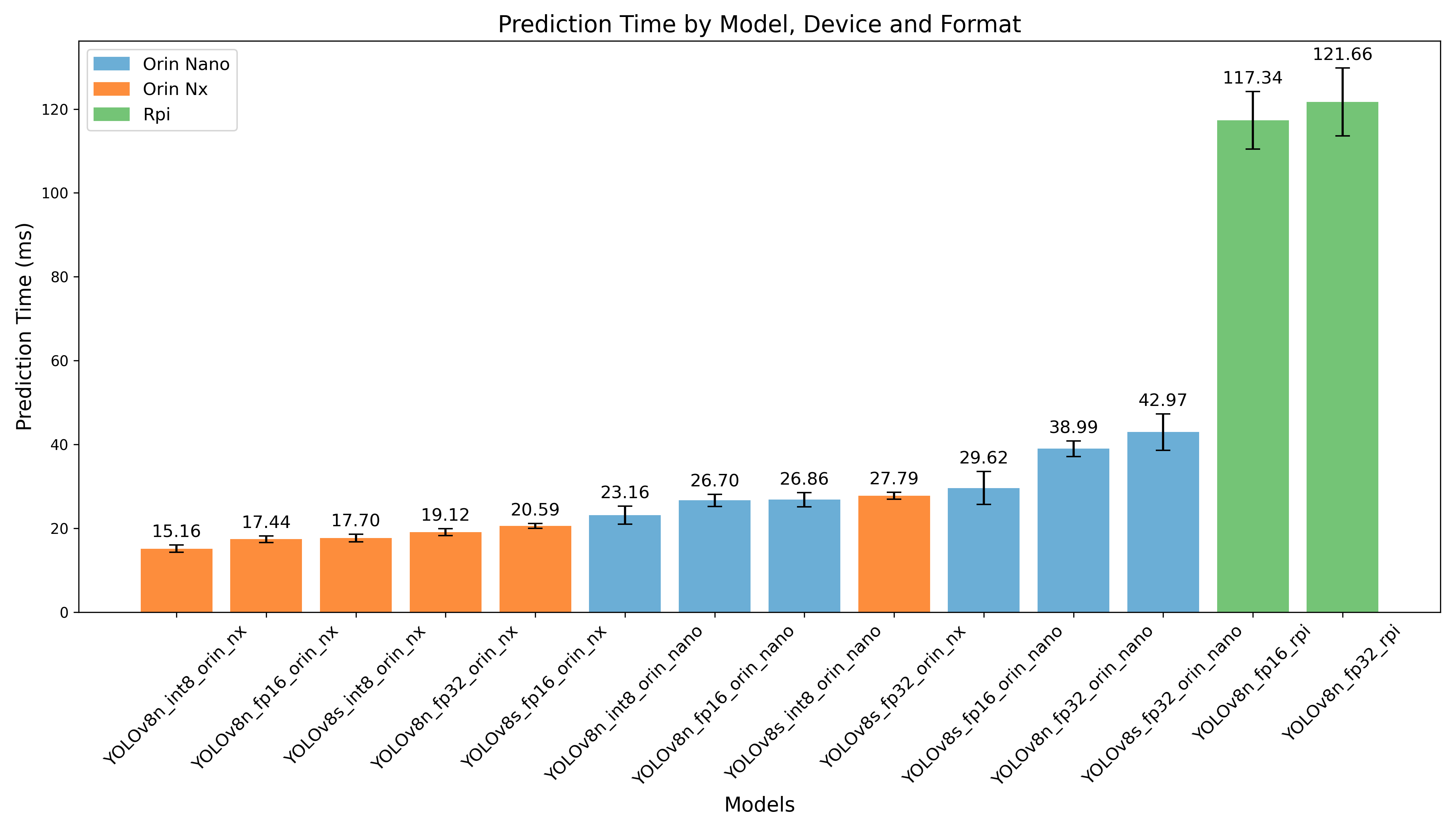}
\end{adjustwidth}
\caption{Figure showing the results on the mean iteration times of each model (YOLOv8s or YOLOv8n) with different quantization versions (FP32, FP16, or INT8) within a device (the Orin NX, Orin Nano, or Raspberry Pi 5).\label{fig:grafica_tiempos}}
\end{figure}

\figurename~\ref{fig:grafica_tiempos} illustrates the Mean Inference Speed (MIS) for the YOLO models on the Orin Nano, Orin NX, and Raspberry Pi devices, comparing the performance with TensorRT (TRT) on the Orin devices and NCNN on the Raspberry Pi.

On the Orin Nano, the YOLOv8n\_INT8 model excelled with an average iteration time of 23.16 ms, followed by YOLOv8n\_FP16 at 26.70 ms and YOLOv8s\_INT8 at \mbox{28.25 ms.} The YOLOv8s\_FP32 model had the longest iteration time of 42.97 ms, illustrating the advantages of INT8 quantization in terms of the time performance. The Orin NX showed an even better performance, with YOLOv8n\_INT8 achieving the shortest iteration time of 15.16 ms, YOLOv8n\_FP16 coming in at 17.44 ms, and YOLOv8s\_INT8 at 17.70 ms.~In contrast, YOLOv8s\_FP32 had the longest time at 27.79 ms, underperforming compared to the leading three Orin Nano models. On the Raspberry Pi, the NCNN-exported models had longer inference times; YOLOv8n\_FP32 averaged at 118.00 ms, and YOLOv8s\_FP32 reached 211.00 ms, highlighting the hardware constraints and differing efficiencies between CPU and GPU model optimizations. In summary, the Orin Nano and Orin NX vastly outperformed the Raspberry Pi, with the Orin NX being suitable for real-time object detection.

During our trials, we attempted to reduce the numerical precision of the YOLOv8 model on the Raspberry Pi 5 using the available libraries. However, neither FP16 nor INT8 configurations were feasible due to limitations in the precision support of the libraries, particularly with the use of the NCNN framework. As a result, there was no improvement in the performance and a notable loss in the prediction accuracy during these tests. Consequently, only the primary FP32 model configuration was included in the comparative analysis, as it was the only stable and functional setup on the Raspberry Pi 5.

\subsection{Degradation Analysis of Model Accuracy}

\begin{figure}[H]
\begin{adjustwidth}{-\extralength}{0cm}
\centering
\includegraphics[width=15.5cm]{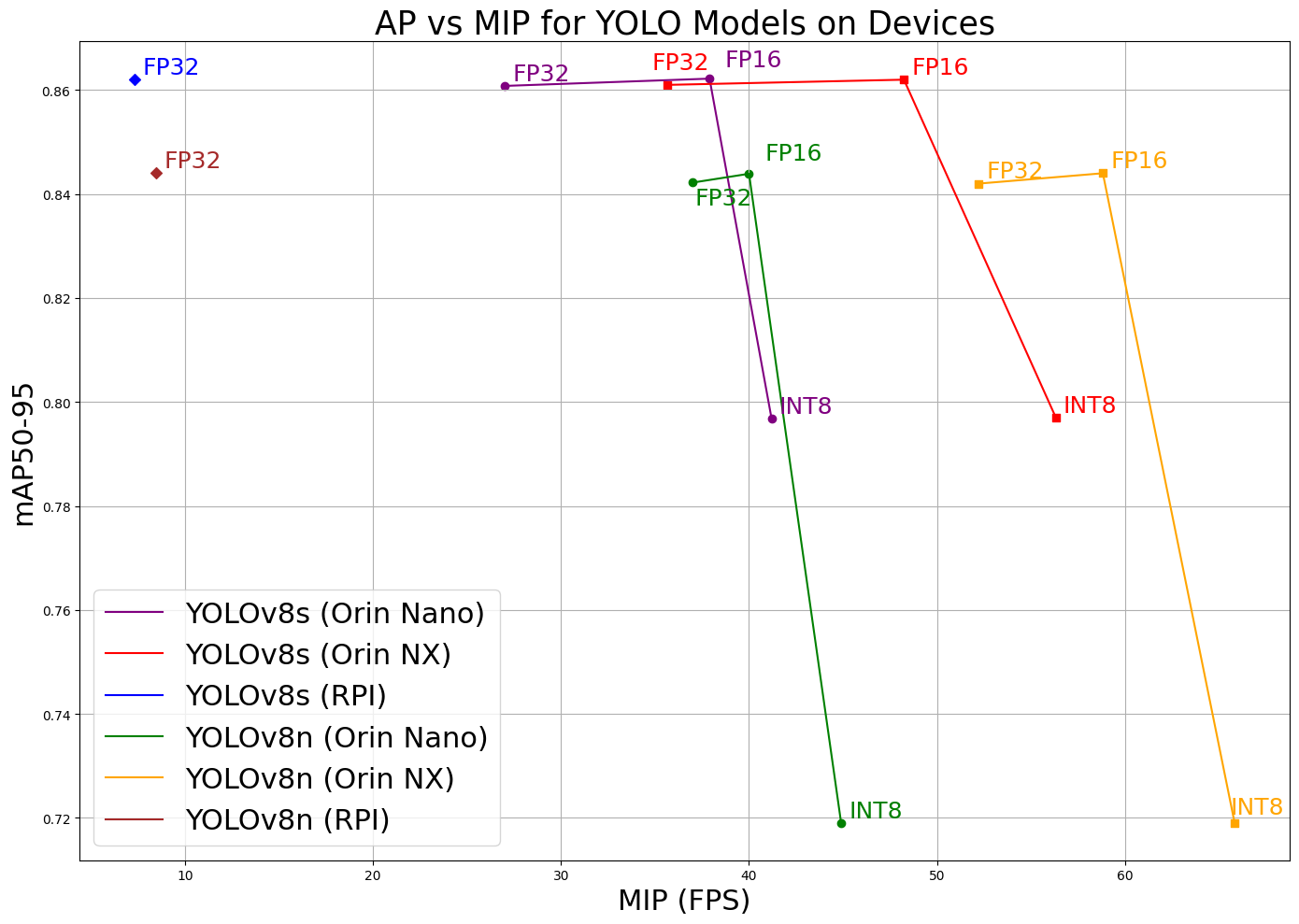}
\end{adjustwidth}
\caption{Figure showing the results of the FPS tests of each isolated model (YOLOv8s or YOLOv8n) with different quantization versions (FP32, FP16, and INT8) within a device.}
\label{fig:mip}
\end{figure}

This section compares the YOLOv8s and YOLOv8n models across various edge devices to assess the impact of quantization on both their inference speed and accuracy. \tablename~\ref{tab:combined} provides a summary of the mAP and FPS results for different devices and quantizations, while \figurename~\ref{fig:mip} illustrates the mean iteration times for each model and configuration.

\begin{table}[H]
\small
\caption{Table synthesizing the inference frame rate, energy consumption, inferences per minute, power, and energy per inference in joules.\label{tab:combined}}
\begin{adjustwidth}{-\extralength}{0cm}
\centering
\renewcommand{\tabcolsep}{0.15cm}
\begin{tabularx}{\fulllength}{CcCccCCCC}
    \toprule
    \textbf{Device} & \textbf{Model} & \textbf{Quantization} & \textbf{mAP50-95} & \textbf{mAP50} & \textbf{MIS (FPS)} & \textbf{ Energy Consumption (W·s)} & \textbf{Inferences per Minute} & \textbf{Energy per Inference (J/inference)} \\
    \midrule
    \multirow[m]{6}{*}{Orin Nano} & YOLOv8s & FP32 & 0.8608 & 0.9771 & 27.00 & 8.790 & 1391 & 0.379 \\ 
                                  & YOLOv8s & FP16 & 0.8622 & 0.9771 & 37.90 & 7.836 & 1558 & 0.302 \\ 
                                  & YOLOv8s & INT8 & 0.7968 & 0.9189 & 41.20 & 7.257 & 1949 & 0.223 \\ 
                                  & YOLOv8n & FP32 & 0.8422 & 0.9619 & 37.00 & 8.345 & 2041 & 0.245 \\ 
                                  & YOLOv8n & FP16 & 0.8439 & 0.9619 & 40.00 & 7.422 & 2144 & 0.208 \\ 
                                  & YOLOv8n & INT8 & 0.7190 & 0.8272 & 44.90 & 7.483 & 2425 & 0.185 \\ 
    \midrule
    \multirow[m]{6}{*}{Orin NX}   & YOLOv8s & FP32 & 0.8610 & 0.9781 & 35.65 & 14.155 & 2129 & 0.399 \\ 
                                  & YOLOv8s & FP16 & 0.8620 & 0.9780 & 48.24 & 12.815 & 2760 & 0.279 \\ 
                                  & YOLOv8s & INT8 & 0.7970 & 0.9195 & 56.32 & 11.002 & 2836 & 0.233 \\ 
                                  & YOLOv8n & FP32 & 0.8420 & 0.9621 & 52.19 & 12.480 & 3182 & 0.235 \\ 
                                  & YOLOv8n & FP16 & 0.8440 & 0.9622 & 58.82 & 10.853 & 3214 & 0.203 \\ 
                                  & YOLOv8n & INT8 & 0.7190 & 0.8276 & 65.83 & 10.305 & 3443 & 0.179 \\ 
    \midrule
    \multirow[m]{2}{*}{RPI5}       & YOLOv8s & FP32 & 0.8620 & 0.9621 & 7.32 & 5.422 & 217 & 1.498 \\ 
                                  & YOLOv8n & FP32 & 0.8440 & 0.9612 & 8.47 & 5.441 & 442 & 0.738 \\ 
    \bottomrule
\end{tabularx}
\end{adjustwidth}
\end{table}

Table~\ref{tab:combined}
presents a summary of the inference performance (in frames per second), energy consumption, the number of inferences per minute, and the energy consumed per inference in joules for the devices Jetson Orin NX, Jetson Orin Nano, and Raspberry Pi 5. The results demonstrate the influence of the disparate quantization configurations (FP32, FP16, INT8) on the energy efficiency and model performance. The energy consumption analysis shows differences between the absolute energy consumption demands and energy efficiency when comparing the Jetson Orin NX, Orin Nano, and Raspberry Pi 5 devices across different quantization methods (FP32, FP16, and INT8). For the Jetson Orin Nano, the energy consumption varies between 7.4 W and 8.7 W, while for the Jetson Orin NX, it ranges from 10 W to 14 W, indicating a more substantial increase.

For the Jetson Orin Nano, the energy consumption ranges from 7.4 W (FP32) to 8.7 W (INT8), reflecting a modest increase of 1.3 W. The Orin NX shows greater variation, ranging from 10 W to 14 W, which aligns with its higher processing capacity. Despite its additional energy demands, the Orin NX delivers notable improvements in its inference speed, particularly with INT8 quantization.

An interesting result is the performance of YOLOv8s on the Orin Nano with FP16 quantization, achieving 37.90 FPS and a mAP50-95 of 0.8622. This configuration slightly exceeds the performance of the Orin NX running YOLOv8s with FP32 (35.65 FPS, mAP50-95 of 0.8610), demonstrating the potential of FP16 quantization to enhance the efficiency even on less advanced hardware. For the YOLOv8n model, the Orin NX achieves 58.82 FPS with FP16, compared to 40.00 FPS on the Orin Nano, indicating its advantage in handling smaller, compact models.

The Raspberry Pi 5, using FP32 quantization via NCNN, achieves 7.32 FPS on YOLOv8s (mAP50-95 of 0.8620). While functional, its performance is limited compared to that of the Jetson devices, suggesting it is better suited to less demanding applications where real-time processing is not critical.

\subsection{Energy Consumption Analysis}
\begin{figure}[H]
    \begin{adjustwidth}{-\extralength}{0cm}
    \centering
    \includegraphics[width=15.5cm]{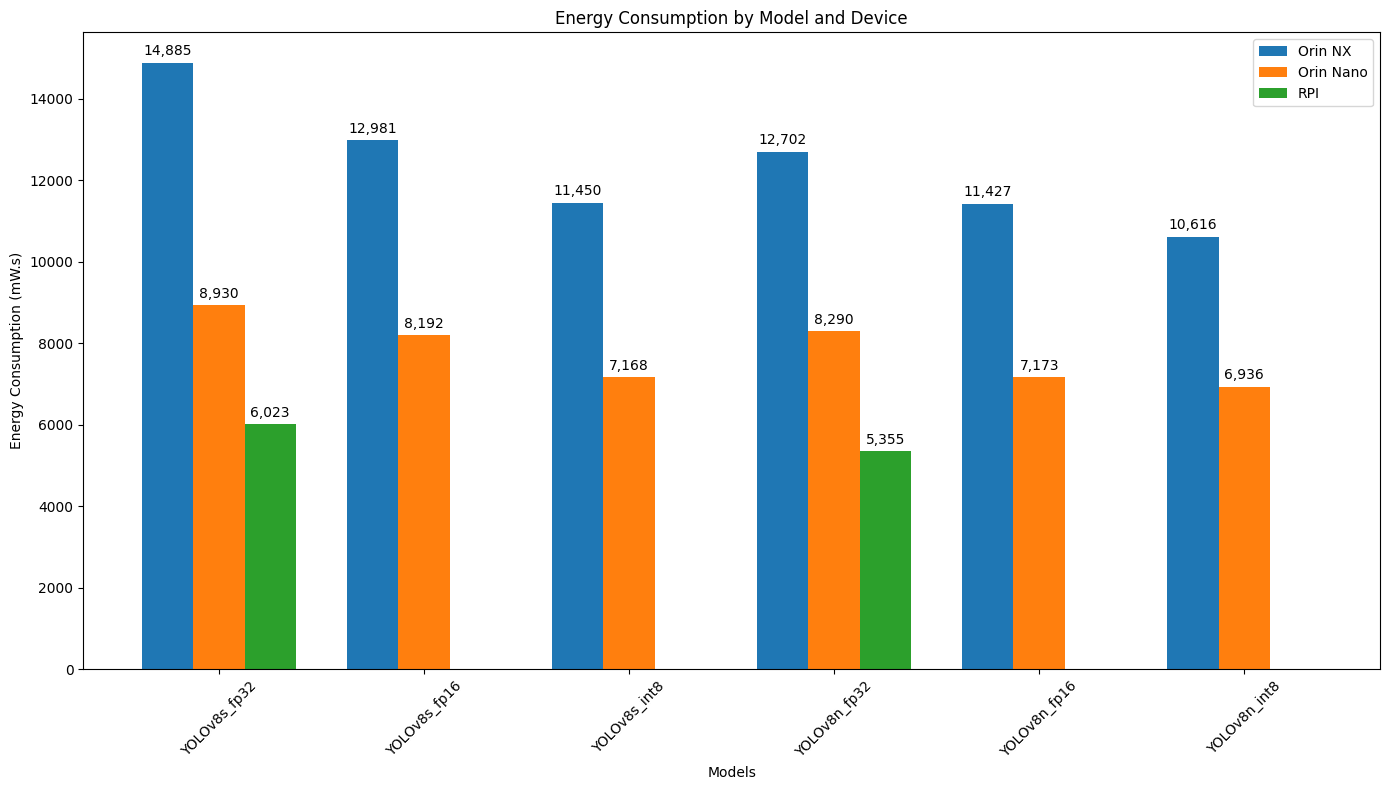}
    \end{adjustwidth}
    \caption{Energy
 consumption by model and device.}
    \label{fig:power_consumption}
\end{figure}

The mean average energy consumption was measured over 5000 images, as depicted in \figurename~\ref{fig:power_consumption}.

 For instance, in the YOLOv8 FP32 setup, the Orin NX consumes 14.155 W, whereas the Orin Nano requires only 8.790 W in the same configuration. Despite the Orin NX’s higher power draw, it generally achieves a greater energy efficiency—measured as the energy per inference (J/inference)—across the configurations due to its higher inference throughput. With INT8 quantization, for example, the Orin NX achieves 0.179 J/inference for YOLOv8n, compared to 0.185 J/inference on the Orin Nano, indicating more efficient use of energy for each inference completed.

The variations in the power consumption between the quantization methods are also noteworthy. For the Orin Nano, the energy consumption rises from 7.4 W in FP32 to 8.7 W in INT8, reflecting a modest increase of 1.3 W. In contrast, the Orin NX exhibits a larger increase, with the power consumption ranging from 10 W in YOLOv8n\_INT8 to 14 W in YOLOv8s\_FP32, a difference of 4 W. 

The Orin Nano demonstrates lower absolute energy consumption and achieves better results in terms of the energy per inference under certain configurations. For instance, in the YOLOv8s\_FP16 setup, it consumes 7.836 W and processes 1558 inferences per minute, resulting in an energy consumption of 0.302 J/inference. Additionally, in FP32 mode, the Orin Nano achieves a better energy efficiency than that of the Orin NX, with 0.379 J/inference compared to 0.399 J/inference.

The Raspberry Pi 5, by comparison, shows the lowest absolute energy consumption but also the lowest energy efficiency due to the small number of predictions per time unit. In the YOLOv8s FP32 configuration, for example, it consumes 5.422 W·s while achieving 217 inferences per minute, resulting in an energy cost of 1.498 J/inference.
The Orin Nano devices, while demonstrating higher absolute energy consumption, also achieves commendable efficiency due to the much better performance in number of inferences per unit of time.

\pagebreak
\section{Performance Evaluation Using a Realistic Testbed}
\label{sec:results2}
\begin{figure}[H]
\begin{adjustwidth}{-\extralength}{0cm}
\centering
\includegraphics[width=15.5 cm]{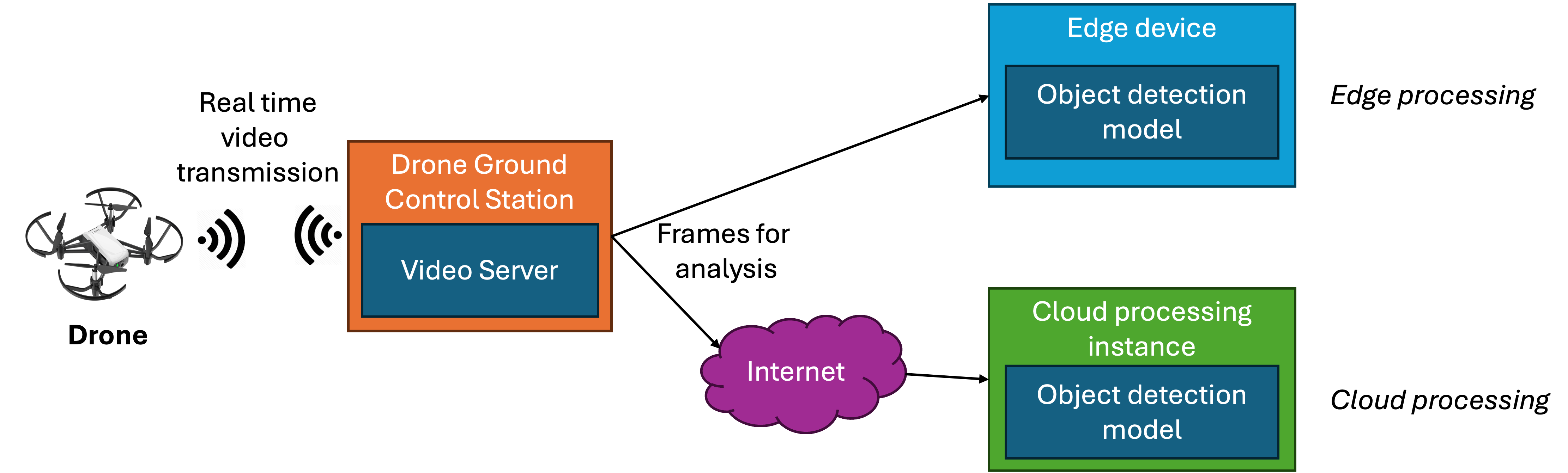}
\end{adjustwidth}
\caption{General system deployment architecture.}
\label{fig:deployment_architecture}
\end{figure}

This section aims to assess the performance of different hardware devices, specifically the Orin Nano, Orin NX, and Raspberry Pi 5, using an object detector in a realistic deployment scenario that mirrors real-world drone operations. In this sense, this section takes advantage of the platform presented by the authors in \cite{David_Carramiñana_IoT} which considers the core elements (e.g., telemetry, video streaming, and data processing) of a real-time drone system, where the interplay of computation and communications affects the latency and system throughput. As depicted in \figurename~\ref{fig:deployment_architecture}, the testbed platform allows live drone images to be processed both in an edge device and in a cloud environment. This enables performance benchmarking of edge algorithms against a centralized cloud, providing a detailed comparison.

\begin{figure}[H]
\begin{adjustwidth}{-\extralength}{0cm}
\centering
\includegraphics[width=15.5 cm]{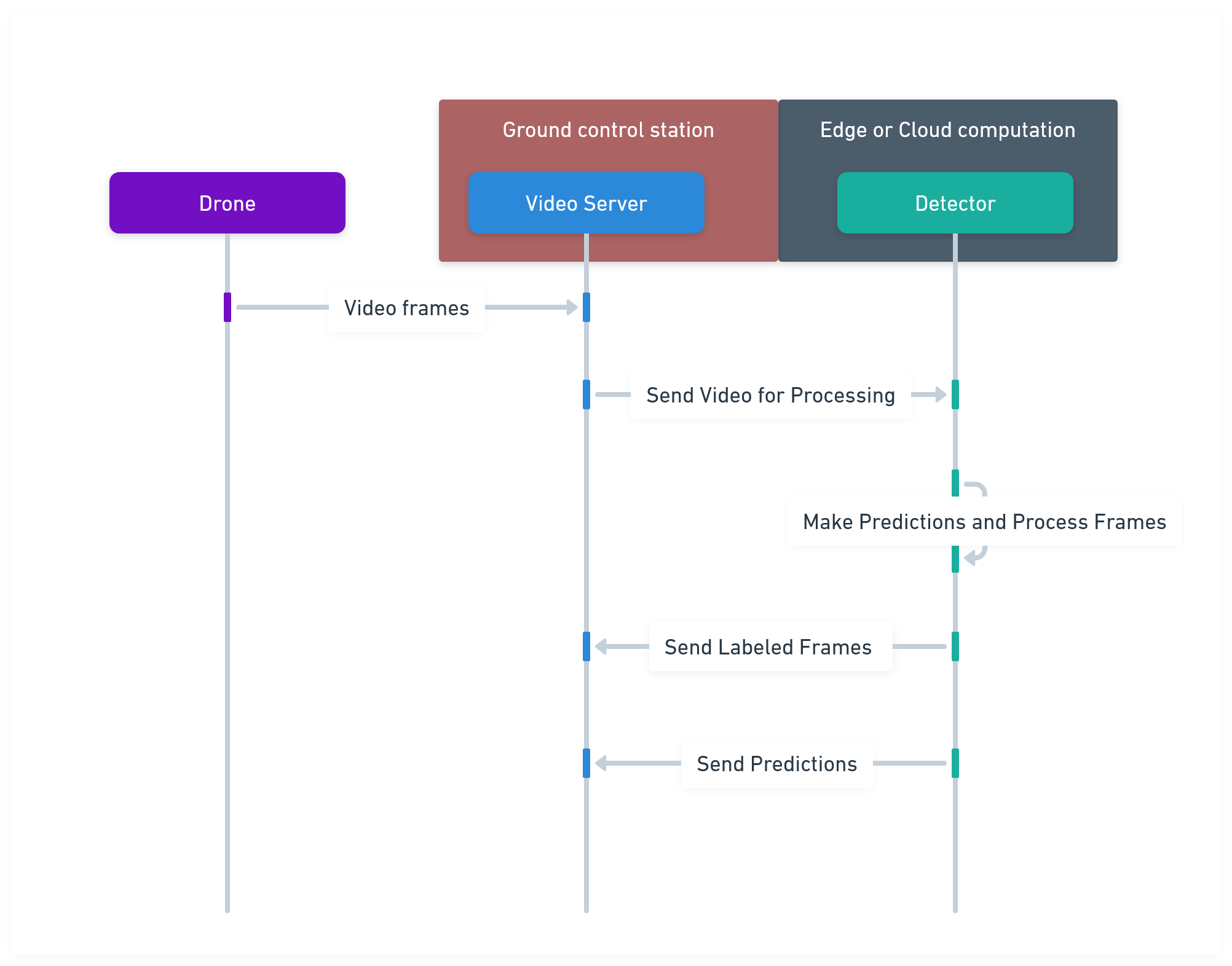}
\end{adjustwidth}
\caption{Data flow for real-time video processing and predictions at the edge.}
\label{fig:fig2}
\end{figure}

The deployment architecture consists of three main components: a drone ground control station (GCS) that gathers video in real time from a drone and broadcasts it using the WebSocket protocol, an edge computing device to perform object detection on video streams using TensorRT optimization, and a cloud processing instance (accessible through the internet) that also allows for the deployment of the object detection model. The interaction between these entities is illustrated in Figure \ref{fig:fig2}. 

The workflow begins with the transmission of video from the drone to its associated ground control station. Then, using WebSocket communication for low-latency and reliable data transfer, the video frames can be forwarded either to an edge computing device, which performs the inference locally (at the edge), or to a cloud instance that performs inference remotely (in the cloud). AWS SageMaker is employed for cloud detection, allowing for a systematic comparison of the latency and processing times between the edge and cloud setups. In either case, the inferences and annotated images are sent back to the ground control station.

Within this framework, two tests were performed. In Section \ref{subsubsec:EdgeDeployment}, different edge devices are compared in the real, representative testbed. Then, in Section \ref{subsubsection:edgevscloud}, a test is performed between edge and cloud processing. The latter evaluation allows for an analysis of the trade-offs between local and centralized processing in terms of the computational power and communication delays. In both cases, by timing each phase (i.e., video transmission, object detection, and prediction distribution), it is possible to identify latency issues and \mbox{potential bottlenecks.}

\subsection{Edge Deployment}
\label{subsubsec:EdgeDeployment}
Table \ref{tab:rtt_throughput} provides a summary of the RTT times, RTT deviations, processing times, and throughput (FPS) for the different models and quantization configurations.

\begin{table}[H]
\small
\caption{Table synthesizing the RTT, RTT std, processing time, and throughput of the system (FPS) for different quantization models.\label{tab:rtt_throughput}}
\begin{adjustwidth}{-\extralength}{0cm}
\centering
\begin{tabularx}{\fulllength}{CCCCCCCC}
    \toprule
    \multirow{2.5}{*}{\textbf{Device}} & \multirow{2.5}{*}{\textbf{Model}} & \multirow{2.5}{*}{\textbf{Quantization}} & \multicolumn{2}{c}{\textbf{RTT (ms)}} & \multicolumn{2}{c}{\textbf{Processing Time (ms)}} & \multirow{2.5}{*}{\shortstack{\textbf{Throughput} \\ \textbf{(inferences/s)}}} \\
    \cmidrule{4-7}
    & & & \textbf{Mean} & \textbf{Std} & \textbf{Mean} & \textbf{Std} & \\
    \midrule
    \multirow[m]{6}{*}{Orin Nano} & YOLOv8n & FP16 & 49.70 & 4.59 & 23.81 & 20.12 & 20.12 \\
                                  & YOLOv8n & FP32 & 60.52 & 12.24 & 28.77 & 16.52 & 16.52 \\
                                  & YOLOv8n & INT8 & 48.80 & 8.42 & 20.54 & 20.49 & 20.49 \\
                                  & YOLOv8s & FP16 & 49.44 & 6.46 & 28.90 & 20.23 & 20.23  \\
                                  & YOLOv8s & FP32 & 73.18 & 4.89 & 47.18 & 13.66 & 13.66 \\
                                  & YOLOv8s & INT8 & 44.27 & 8.27 & 20.62 & 22.59 & 22.59 \\
    \midrule
    \multirow[m]{6}{*}{Orin NX}   & YOLOv8n & FP16 & 34.15 & 3.65 & 16.12 & 29.28 & 29.28 \\
                                  & YOLOv8n & FP32 & 35.76 & 2.28 & 18.25 & 27.97 & 27.96 \\
                                  & YOLOv8n & INT8 & 29.95 & 3.19 & 14.29 & 33.38 & 33.39 \\
                                  & YOLOv8s & FP16 & 35.09 & 4.35 & 18.50 & 28.50 & 28.5 \\
                                  & YOLOv8s & FP32 & 48.97 & 8.83 & 26.50 & 20.42 & 20.42 \\
                                  & YOLOv8s & INT8 & 33.60 & 3.31 & 15.96 & 29.76 & 29.76 \\
    \bottomrule
\end{tabularx}
\end{adjustwidth}
\end{table}

Regarding the RTT (Round-Trip Time), the Jetson Orin Nano demonstrated its best performance in terms of the RTT with the YOLOv8s INT8 model, achieving a minimum RTT of 44.27 ms. The RTT increased to 73.18 ms when using the YOLOv8s FP32 model. For the Jetson Orin NX, the minimum RTT was observed at 29.95 ms with the YOLOv8n INT8 model, while the maximum RTT occurred with the YOLOv8s FP32 model at 48.97 ms.

The RTT deviation results showed that the Orin Nano showed higher variability with the YOLOv8n FP32 model, which had a deviation of 12.24 ms. The lowest deviation for the Orin Nano was seen with the YOLOv8s FP32 model at 4.89 ms. For the Orin NX, the RTT fluctuations were smaller overall, with the highest deviation being 8.83 ms for the YOLOv8s FP32 model and the lowest at 2.28 ms for YOLOv8n FP32.

During the testing phase, we attempted to implement the architecture on the Raspberry Pi 5. However, it was not possible to complete the tests in full due to the computational limitations. This device exhibited significant latency in its predictions, and this led to instability not only in drone control but also in the acquisition and dissemination of images across the different components of the architecture. Consequently, the results obtained from the Raspberry Pi 5 are sparse and incomplete and therefore not viable as a reliable architecture for this application.

In terms of the processing time, the Orin Nano showed a minimum processing time of 20.54 ms with the YOLOv8n INT8, and the longest processing time was 47.18 ms for the YOLOv8s FP32 model. The processing times on the Orin NX were generally shorter, with a minimum of 14.29 ms for YOLOv8n INT8 and a maximum of 26.50 ms for YOLOv8s FP32.

\subsection{Edge vs. Cloud Comparison}
\label{subsubsection:edgevscloud}
In order to assess the trade-offs between computational power and latency in 
edge and cloud environments, we conducted a series of experiments comparing the two setups using the YOLOv8s model in its FP16 configuration on the Orin Nano. The primary metrics evaluated were the Round-Trip Time (RTT), model processing latency, and communication latency, offering a comprehensive understanding of how both environments handle real-time data processing and communication delays.

For our cloud-based deployment, we selected a cloud instance located in London (eu-west-2) using the \textit{ml.g4dn.xlarge}
configuration, which is powered by an NVIDIA T4 GPU. To handle the model inferences efficiently, we deployed the model on a Triton Inference Server within the SageMaker environment. Triton, known for its high performance and flexibility in supporting multiple AI frameworks and backends, provided a robust platform for managing the inference requests for our YOLOv8s model. The most decisive factor in choosing the NVIDIA Triton Inference Server was its support for TensorRT optimization, which allowed us to conduct a fair comparison under consistent conditions. Additionally, by utilizing Triton, we could take advantage of advanced features such as dynamic batching, model ensemble, and TensorRT acceleration, thereby maximizing the inference speed and fully leveraging the computational power of the cloud instance.

The model was deployed and invoked directly through the SageMaker SDK API's invoke endpoint, which facilitated direct HTTP-based invocation. However, it is important to note that the invoke endpoint does not support WebSocket communication, meaning that all inference requests are handled over HTTP. This limitation may introduce additional latency in scenarios where persistent, real-time communication channels are desired.
 The London (eu-west-2) location helped us explore the impact of geographical data transmission delays from Madrid, providing insights into latency management for future applications.

Table \ref{tab:cloud-results} presents a comparative investigation of the latency metrics between the edge deployments on the Orin Nano and the cloud deployments using the YOLOv8s model in FP16 format. The edge RTT latency shows a mean of 35.09 ms, which is within the feasible limits for real-time applications. By contrast, the cloud deployment shows a much higher mean RTT of 348.21 ms due to the increased communication distance to remote GPU servers. This difference is reflected in the communication latency, where the edge setup averages at 2.50 ms, whereas the cloud averages at a considerably higher value of 341.41 ms.

\begin{table}[H]
\caption{Latency results in the two different deployment scenarios, edge and cloud.\label{tab:rtt_latency}}\label{tab:cloud-results}
\begin{adjustwidth}{-\extralength}{0cm}
\centering
\begin{tabularx}{\fulllength}{CCCC}
    \toprule
    \textbf{Processing Environment} & \textbf{RTT Latency (ms)} & \textbf{Model Processing Latency (ms)} & \textbf{Communication Latency (ms)} \\
    \midrule
    \multirow{2}{*}{Edge} & Mean: 35.09 & Mean: 32.59 & Mean: 2.50 \\
                          & Std: 4.25  & Std: 4.27  & Std: 0.34  \\
    \midrule
    \multirow{2}{*}{Cloud} & Mean: 348.21 & Mean: 6.82  & Mean: 341.41 \\
                           & Std: 69.88   & Std: 0.05  & Std: 69.89  \\
    \bottomrule
\end{tabularx}
\end{adjustwidth}
\end{table}

The processing latency data highlight the computational edge of the cloud, with an average latency of 6.82 ms, significantly outperforming the edge's latency of 32.59 ms. Nonetheless, the high communication latency of the cloud counteracts its processing efficiency, resulting in performance constraints.

\section{Discussion}
\label{sec:discussion}
An evaluation of various quantization techniques and their impact on inference latency has confirmed INT8 as the fastest configuration, achieving notable reductions in the processing times across all evaluated devices. For instance, on the Orin Nano, YOLOv8s with INT8 achieves 41.20 FPS compared to 27.00 FPS with FP32, representing a significant improvement in speed. However, this increase comes at the cost of reduced accuracy, with the mAP50-95 dropping from 0.8608 in FP32 to 0.7968 in INT8—a reduction of approximately 6.4 percentage points. In contrast, FP16 emerges as a balanced alternative, offering slight improvements in accuracy while reducing the latency. For example, YOLOv8s on the Orin Nano achieves 37.90 FPS in FP16 with an accuracy of 0.8622 mAP50-95, providing an optimal balance for scenarios requiring both speed and precision. This quantization approach also enables intermediate models, such as YOLOv8s, to outperform smaller versions like YOLOv8n\_FP32 on more advanced devices such as the Orin NX, where YOLOv8s\_FP16 achieves 48.24 FPS with a mAP50-95 of 0.8620, compared to YOLOv8n\_FP32's 35.65 FPS and 0.8420 mAP50-95. Furthermore, the application of quantization techniques has been observed to contribute to marginal enhancements in the model's generalization, possibly due to the reduced computational complexity, enabling more efficient image processing. This is particularly relevant for tasks involving real-time object detection, where the computational resources are constrained. Conversely, while the Raspberry Pi 5 demonstrates acceptable accuracy metrics, such as a mAP50-95 of 0.8620 for YOLOv8s in FP32, its inference latency limits its applicability. The device achieves only 7.32 FPS and 217 inferences per minute in this configuration, significantly reducing the real-time capabilities.
In addition to optimizing performance, quantization techniques have a significant impact on the energy efficiency of employing devices. The results demonstrate that INT8 is the most efficient configuration in terms of energy consumption, achieving a notable reduction without a significant loss of accuracy. In this context, the Orin NX, despite its higher overall power demand, demonstrates a better performance in the EPI metric (J/inference). This makes it the preferred choice for real-time, inference-intensive applications, especially in the FP16 and INT8 configurations. The Raspberry Pi 5 has the lowest total power consumption; however, its low performance in terms of Inferences Per Minute (IPM) restricts its applicability in environments where a real-time response is required. This analysis highlights the importance of selecting optimized hardware devices, such as the Orin models, for edge architectures that demand both high performance and sustainable energy efficiency, particularly in long-running applications where controlled power consumption is essential.

The evaluation of the stability and performance of the devices within architectures that emulate real-world operating environments highlights the advantage of the Orin models over general-purpose alternatives such as the Raspberry Pi 5. In practical scenarios, the models evaluated on the Orin Nano achieved acceptable results; however, a detailed analysis suggests that improvements can still be made in frame capture and the optimization of internal operations, which could reduce the processing overhead. This overhead, identified by analyzing isolated model processing times during the full operation of the architecture, suggests that better management of data capture and data flow could significantly \mbox{improve the efficiency.}

The Orin NX exhibited low variability in its response times and strong stability, positioning it as optimal for tasks demanding a consistent performance throughout the architecture. Conversely, the Raspberry Pi 5 encountered issues with instability, processing interruptions, and control challenges during object detection, constraining its effectiveness in complex architectures. Thus, it is more appropriate for basic processing or simpler algorithms rather than advanced applications requiring sustained high stability and low latency.

In this way, the results reveal complementary strengths in the tested devices: the Jetson excels in its computational performance but consumes significant energy, while the Raspberry Pi offers excellent energy efficiency but struggles with real-time object detection. Hybrid architectures could combine these strengths to optimize the overall performance. For example, the Raspberry Pi could manage lightweight tasks, such as environmental monitoring or data preprocessing, while the Jetson handles critical operations like real-time object detection or obstacle avoidance. This delegation of tasks reduces the energy footprint of the individual components while maintaining the system's responsiveness. Additionally, selectively integrating cloud processing for non-critical tasks enhances the scalability and reduces latency. Exploring this approach in future implementations could help address challenges in energy-sensitive drone missions.

The cloud results underscore the advantage of deploying edge devices near data sources. Despite the low model processing latency in the cloud, the high communication latency limits its real-time viability. Using a closer AWS region may improve the results, but communication bottlenecks remain a challenge for real-time drone applications. GPU-equipped edge devices offer a practical solution, enabling rapid local processing with some trade-offs in the computational power, showing the importance of the role of edge devices in real-time drone systems.

Based on the results obtained and from the perspective of this study, focused on drone applications, a guide has been developed to identify the most suitable hardware configurations and quantization models according to the operational needs. This guide, based on the lessons learned, provides recommendations for selecting devices and model configurations based on the specific requirements, such as extended coverage, real-time tracking, and \mbox{high-precision detection.} 

Table~\ref{tab:recommended_configurations} illustrates how particular configurations align with specific requirements in drone applications. For extended coverage and long-duration missions, the Orin NX with INT8 offers the optimal autonomy with minimal inference consumption, making it ideal for continuous surveillance and environmental monitoring. In critical tracking scenarios, the same device with YOLOv8n INT8 enables a high inference rate, which is valuable for real-time tasks in defense and security. The Orin Nano and FP16 quantization enable high-quality detection, which is a requirement in automated infrastructure inspection, while maintaining the optimal power consumption. The INT8 configuration of the Orin Nano allows low-demand tasks to be completed efficiently and cost-effectively, even when speed or accuracy is not critical. These results demonstrate the value of aligning the configuration with specific operational demands, ensuring that each device performs at its best in its designated application. These insights serve as practical guidance for drone operators and practitioners, helping them select the most appropriate hardware and quantization settings based on the specific operational needs. Researchers should examine scalability in more UAV scenarios (e.g., evaluating the outdoor performance variability or investigating energy-efficient inference to provide insights and new extended guidelines for real-world deployment). Moreover, adapting the results obtained in the indoor controlled edge environment to outdoor surroundings would assess the stability of inferences under dynamic conditions, thereby verifying the practicality of energy-efficient setups outside controlled contexts. Furthermore, as the industry advances in the hardware--software co-design for FPGAs and low-power chips, research should incorporate these technologies to align with \mbox{emerging capabilities.}

\begin{table}[H]
\caption{Recommended configurations based on operational needs.\label{tab:recommended_configurations}}
\begin{adjustwidth}{-\extralength}{0cm}
\centering
\begin{tabularx}{\fulllength}{CCCCC}
    \toprule
    \textbf{Operational Need} & \textbf{Description} & \textbf{Area of Interest} & \textbf{Recommended Configuration} & \textbf{Key Supporting Metrics} \\
    \midrule
    Extensive Coverage and Long Duration & Extends energy autonomy for continuous monitoring of large areas without frequent recharging. & Surveillance, Environmental Monitoring & Jetson Orin Nano + YOLOv8n INT8 & High FPS (44.9), Low Energy Consumption (\mbox{7.483 W·s}), Good Accuracy (mAP50: 0.8272) \\
    \midrule
    Critical Tracking (Real-Time) & High inference rate for rapid response and continuous tracking. & Defense, Security & Jetson Orin NX + YOLOv8n INT8 & Very High FPS (65.83), Moderate Energy (10.305 W·s), Sufficient Accuracy (mAP50: 0.7190) \\
    \midrule
    High-Quality Detection & Ensures high precision in detecting fine details, ideal for infrastructure inspection. & Infrastructure Inspection & Jetson Orin Nano + YOLOv8s FP16 & High Accuracy (mAP50: 0.8622), Moderate FPS (37.9), Low Energy Consumption (\mbox{7.836 W·s}) \\
    \midrule
    Balance of Speed and Precision & Balances speed and accuracy for emergency and rescue tasks. & Rescue and Emergency Response & Jetson Orin NX + YOLOv8s FP16/YOLOv8n INT8 & High FPS (58.82/65.83), Balanced Accuracy (mAP50: 0.8620/0.7190), Efficient Energy Usage (\mbox{10.853/10.305 W·s}) \\
    \midrule
    Low-Cost Processing & Optimizes costs and energy for less demanding tasks without high precision or speed requirements. & Low-Demand Tasks & Jetson Orin Nano + INT8 & Moderate FPS (41.2), Low Energy Consumption (\mbox{7.257 W·s}), Acceptable Accuracy (mAP50: 0.7968) \\
    \bottomrule
\end{tabularx}
\end{adjustwidth}
\end{table}

\section{Conclusions}
\label{sec:conclusions}

This research provides practical tools and guidelines for integrating YOLOv8 models into autonomous systems while also evaluating their performance on computationally constrained drone computational devices. The findings show that the YOLOv8n (nano) and YOLOv8s (small) models can achieve efficient operation on devices such as the Raspberry Pi 5, Orin NX, and Jetson Orin Nano. However, achieving the optimal performance in real-time applications requires careful consideration of the trade-offs between the detection precision and inference speed, especially when utilizing INT8 quantization, which accelerates the processing but may slightly reduce the accuracy.

Thus, we provide a set of recommendations (\tablename~\ref{sec:discussion}) that streamlines complex recommendations into an easily accessible format for decision-makers to use to adjust their configurations according to the specific operational demands.
 By linking the device configurations and quantization methods with specific applications, this study explores the practical application of guidelines that enhance the performance and energy efficiency. This organized method clarifies the criteria for the optimal configuration selection while enabling practitioners to efficiently implement these advances in diverse edge computing scenarios.

The test architecture in this study facilitates controlled and repeatable testing of real-time object detection and tracking algorithms.~It enables an accurate evaluation of edge computing and onboard processing solutions within a comprehensive drone architecture in an indoor flight environment. However, the indoor setting introduces certain limitations, as it cannot fully replicate the complexities of outdoor operations, such as variable lighting conditions, environmental interferences, or multi-agent dynamics. To address these challenges, more complex mock-ups are needed to emulate more complex environments. Such improvements would enable testing under conditions closer to those in real-world applications, improving the reliability of the proposed systems.

An interesting direction for upcoming research is the evaluation of YOLOv10 and YOLOv11 as they become available, evaluating their improvements in accuracy, efficiency, and real-time performance. Comparing them with YOLOv8 in drone-based applications could provide insights into their suitability for edge computing and its integration viability in autonomous drone navigation.

Moreover, studies should explore hybrid architectures that combine the complementary strengths of devices like the Jetson, the Raspberry Pi, and the cloud. As mentioned in \mbox{Section \ref{sec:discussion},} Discussion, its complementary abilities may offer a study opportunity. Testing these configurations in real-world drone deployments would provide valuable insights into their feasibility and effectiveness in energy-sensitive missions.

Further integration of edge architectures with 5G and tactical cloud technologies could expand the potential of drone applications. This integration would enable the design and control of complex swarm missions, as discussed in Section \ref{sec:sota}, The State of the Art. Through network slicing, the processing tasks could be decentralized, increasing the autonomy of individual drones within the swarm and reducing reliance on a central server. Moreover, testing 5G communications in these configurations could provide insights into intelligent drone architectures, particularly regarding real-time decision-making and minimizing the latency in data transmission. These advancements would also necessitate the deployment of larger drones capable of supporting the additional weight and power requirements associated with outdoor operations in complex environments.

\vspace{6pt}

\authorcontributions{Conceptualization: A.M.B. Methodology: A.M.B. and D.C. Software: L.R. and A.D.D. Validation: L.R., D.C., and A.D.D. Formal analysis: J.A.B., L.B., and L.R. Investigation: L.R. and D.C. Data curation: L.R. and A.D.D. Resources: L.B. and J.A.B. Writing---original draft preparation: L.R., A.M.B., A.D.D., and L.B. Writing---review and editing: all. Visualization: L.R. Supervision: A.M.B., J.A.B., and J.R.C. Project administration: A.M.B. Funding acquisition: J.R.C. and A.M.B. All of the authors have read and agreed to the published version of the manuscript.}

\funding{This research was funded by the Horizon Europe EDF Programme grant number 101103386 and under the national grants TSI-063000-2021-80 and MIA.2021. M04.0008 by the Spanish Ministry of Economic Affairs and Digital Transformation and by the EU Next Generation EU/PRTR programme. MCIN/AEI/10.13039/501100011033 under Grant PID2020-118249RB-C21.}

%

\dataavailability{The raw data supporting the conclusions of this article will be made available by the authors on request.}

\conflictsofinterest{The authors declare no conflicts of interest.} 

\begin{adjustwidth}{-\extralength}{0cm}

\reftitle{References}



\bibliographystyle{mdpi}


%


\PublishersNote{}
\end{adjustwidth}
\end{document}